\documentclass[twocolumn,prb,aps]{revtex4}
\usepackage{graphicx}
\usepackage{dcolumn}
\usepackage{amsmath}
\usepackage{color}
\newcommand{\be}{\begin{equation}}
\newcommand{\ee}{\end{equation}}
\newcommand{\bea}{\begin{eqnarray}}
\newcommand{\eea}{\end{eqnarray}}
\newcommand{\bs}{\begin{split}}
\newcommand{\bse}{\begin{subequations}}
\newcommand{\ese}{\end{subequations}}

\newcommand{\lumnsi}{Lu$_{1-x}$Sc$_{x}$MnSi}
\newcommand{\degr}{$^\circ$}
\providecommand{\e}[1]{\ensuremath{\times 10^{#1}}} 

\begin{document}
\title{Helical Antiferromagnetic Ordering in \lumnsi\ ($x=$ 0, 0.25, 0.50)}
\author {R. J. Goetsch}
\author {V. K. Anand}
\altaffiliation{Present address: Helmholtz-Zentrum Berlin f\"ur Materialien und Energie, Hahn-Meitner Platz~1, 14109 Berlin, \mbox{Germany}.}
\author {D. C. Johnston}
\altaffiliation{johnston@ameslab.gov}
\affiliation {Ames Laboratory and Department of Physics and Astronomy, Iowa State University, Ames, Iowa 50011, USA}


\date{\today}

\begin{abstract}

Polycrystalline samples of \lumnsi\ ($x=$ 0, 0.25, 0.50) are studied using powder x-ray diffraction (XRD), heat capacity $C_{\rm p}$, magnetization~$M$, magnetic susceptibility $\chi$, and electrical resistivity $\rho$ measurements versus temperature~$T$ and magnetic field~$H$\@. This system crystallizes in the primitive orthorhombic  TiNiSi-type structure (space group {\it Pnma}) as previously reported. The $\rho(T)$ data indicate metallic behavior.  The $C_{\rm p}(T)$, $\chi(T)$, and $\rho(T)$ measurements consistently indicate long-range antiferromagnetic (AF) transitions with AF ordering temperatures $T_{\rm N} = 246$, 215 and 188~K for $x=0$, 0.025 and 0.50, respectively.  A second transition is observed at somewhat lower~$T$ for each sample from the $\chi(T)$ and $\rho(T)$ measurements, which we speculate are due to spin reorientation transitions; these second transitions are completely suppressed in  $H=5.5$~T\@. The $C_{\rm p}$ data below 10~K for each composition indicate an enhanced Sommerfeld electronic heat capacity coefficient for the series in the range $\gamma =$~24--29~mJ/mol\,K$^2$.  The $\chi(T)$ measurements up to 1000~K were fitted by local-moment Curie-Weiss behaviors which indicate a low Mn spin $S\sim1$. The $\chi$ data below $T_{\rm N}$ are analyzed using the Weiss molecular field theory for a planar noncollinear helical AF structure with a composition-dependent pitch, following the previous neutron diffraction work of Venturini et al.~[J. Alloys Compd. {\bf 256}, 65 (1997)]. Within this model, the fits indicate a turn angle between Mn moments along the helix axis of $\sim 100^\circ$ or $\sim 145^\circ$, either of which indicate dominant AF interactions between the Mn spins in the \lumnsi\ series of compounds.

\end{abstract}


\maketitle

\section{Introduction}

Competing magnetic interactions often lead to noncollinear magnetic structures,\cite{Diep2004} such as occurs in triangular lattice antiferromagnets (AFs) with the famous 120\degr\ ordering.\cite{Collins1997} Such materials are interesting in their own right but a noncollinar magnetic structure can also occur in conjunction with other ordered states such as in  magnetoelectric multiferroics.  In one class of magnetoelectrics, the coupling between the electric and magnetic degrees of freedom requires a helical AF spin structure.\cite{Tokura2010}  The noncollinear magnetic structures of such materials are usually determined using neutron diffraction measurements.

One of us (DCJ) recently formulated a molecular field theory (MFT) of the anisotropic magnetic susceptibility $\chi$ below the N\'eel temperature $T_{\rm N}$ of single-crystal planar noncollinear Heisenberg AFs containing identical crystallographically-equivalent spins.\cite{Johnston2012}  This MFT allows details of the magnetic structure as well as of the magnetic interactions to be estimated from a fit to the $\chi$ versus temperature $T$ data in the AF state at $T < T_{\rm N}$.  The predictions are useful because they are generic, complementary to neutron diffraction measurments, and are applicable to a variety of noncollinear AF structures and materials.  MFT does not account for quantum fluctuations, so it is expected to be most accurate for three-dimensional spin lattices with large spin~$S$\@.  On the other hand, deviations of experimental $\chi(T<T_{\rm N})$ data from the theory can be used as a diagnostic for spin fluctuations and correlations beyond MFT, as already illustrated by fits to the anisotropic $\chi(T<T_{\rm N})$ for several materials.\cite{Johnston2012}  The MFT also makes the remarkable prediction that the triangular lattice AF should have an {\it isotropic and $T$- and $S$-independent} $\chi(T)$ at $T < T_{\rm N}$, as has actually been observed for many such antiferromagnets.\cite{Collins1997, Johnston2012}  Although the MFT was formulated to fit the anisotropic $\chi(T)$ of single crystals, it is used in this paper to model $\chi(T\leq T_{\rm N})$ data for polycrystalline samples by taking the powder average of the single-crystal predictions.

The above MFT has mostly been applied to fitting the anisotropic $\chi(T)$ for single-crystal AF insulators,\cite{Johnston2012} with the exception of a polycrystalline sample of the stoichiometric metallic compound ${\rm Y_3MnAu_5}$, where a helical AF structure with a turn angle of 69\degr\ was predicted from the fit to $\chi(T\leq T_{\rm N})$.\cite{Samal2012}  A figure showing this helical structure is shown in Fig.~1 of Ref.~\onlinecite{Johnston2012}, which consists of ferromagnetically-aligned layers of spins in the $xy$ plane, with the ordered moments aligned within the layer, where the direction of the ordered moments rotates by a fixed angle between 0 and 180\degr\ from layer to layer along the $z$ axis.

The main goal of the present work is to apply the MFT to fit $\chi(T<T_{\rm N})$ for polycrystalline samples of the metallic \lumnsi\ system which crystallizes in the orthorhombic TiNiSi-type structure.\cite{Venturini1997, Ijjaali1999} Neutron diffraction measurements by Venturini et al.\ indicated that this system exhibits a proper-helix AF ground state for $x = 0$ and $x=0.9$,\cite{Venturini1997} where the magnitude of the pitch of the helix depends on the Sc concentration~$x$. Since our MFT formulation predicts that $\chi(T<T_{\rm N})$ is sensitive to the pitch of a helical AF structure,\cite{Johnston2012} this system is attractive for applying and testing the MFT predictions.

Here we present x-ray diffraction (XRD), magnetization $M$ versus applied magnetic field $H$, $\chi(T)$, heat capacity $C_{\rm p}(T)$ and electrical resistivity $\rho(T)$ measurements on polycrystalline samples of \lumnsi\ with compositions $x=0$, 0.25 and 0.50.  Although some of our data suggest that \lumnsi\ may be an itinerant AF, itinerant AFs can often be parameterized by a local-moment Heisenberg model as was done for iron-arsenide high-$T_{\rm c}$ superconductors and parent compounds.\cite{Johnston2010} 

We present the experimental details in Section~\ref{ExpDetails} and our crystallography study is described in Sec.~\ref{Structure}.  The $M(H)$ and $\chi(T)$, $C_{\rm p}(T)$, and $\rho(T)$ measurements are presented and analyzed in Secs.~\ref{Transport}--\ref{HC}, followed by a summary in Sec.~\ref{Conclusion}.

\section{\label{ExpDetails} EXPERIMENTAL DETAILS}

Polycrystalline samples of \lumnsi\ ($x=0$, 0.25, and 0.50) were prepared by arc-melting the high purity elements Lu from Ames Laboratory, Mn (99.99\%) from Alfa Aesar and Si (99.999995\%) from ROC/RIC under ultra high purity argon.  The samples were turned over and remelted several times to ensure homogeneity. Fifteen wt\% extra Mn was included in the starting composition to account for mass loss during arc-melting. The arc-melted samples were then annealed at 800~\degr C for one week.  As evident in the powder XRD patterns presented in Fig.~\ref{fig:ScDoped_XRD} below, the samples are nearly single phase.  We attempted to prepare samples with higher doping levels than $x=0.5$, but these samples were two-phase with large amounts of ScMnSi which forms with a different structure.  In addition, unsuccessful attempts to grow single crystals with the correct phase were attempted from solution using Sn, Al, Cu, and Ga as fluxes.

Powder XRD patterns were obtained with a Rigaku Giegerflex powder diffractometer using Cu~K$_\alpha$ radiation and the crystal structures were determined using Rietveld profile analysis with the {\tt FullProf} package.\cite{Rodriguez1993}

The $M(H,T)$ and $\chi(T)$ measurements from 1.8 to 300~K were carried out using a Quantum Design, Inc., magnetic properties measurement system (MPMS). A gelatin capsule was used as a sample holder in these measurements. For high-temperature magnetization measurements up to 1000~K, the vibrating sample magnetometer (VSM) option of the Quantum Design, Inc., physical properties measurement system (PPMS) was used. In all magnetic measurements, the sample holder was measured separately and corrected for in the data presented here. 

The $C_{\rm p}(T)$ measurements were done using a Quantum Design PPMS\@. The samples had masses of 10--15~mg and were thermally anchored to the heat capacity addenda using Apiezon N~grease.

The $\rho(T)$ was measured using the ac~transport option of the PPMS\@. These measurements utilized rectangular-shaped samples that were cut from the arc-melted buttons using a low-speed diamond wheel saw.  Platinum electrical leads were attached to the samples using silver epoxy. These measurements were performed on both cooling and heating to check for thermal hysteresis; however, no hysteresis was detected for any of the samples, indicating that the magnetic transitions in the samples are thermodynamically of second order, consistent with the results of the $C_{\rm p}(T)$ measurements.  Due to uncertainties in the geometric factor, the magnitude of the $\rho$ data have a systematic error of order~10\%. 

\section{\label{Structure} Crystallography}
\begin{figure}
	\includegraphics[width=2.7in]{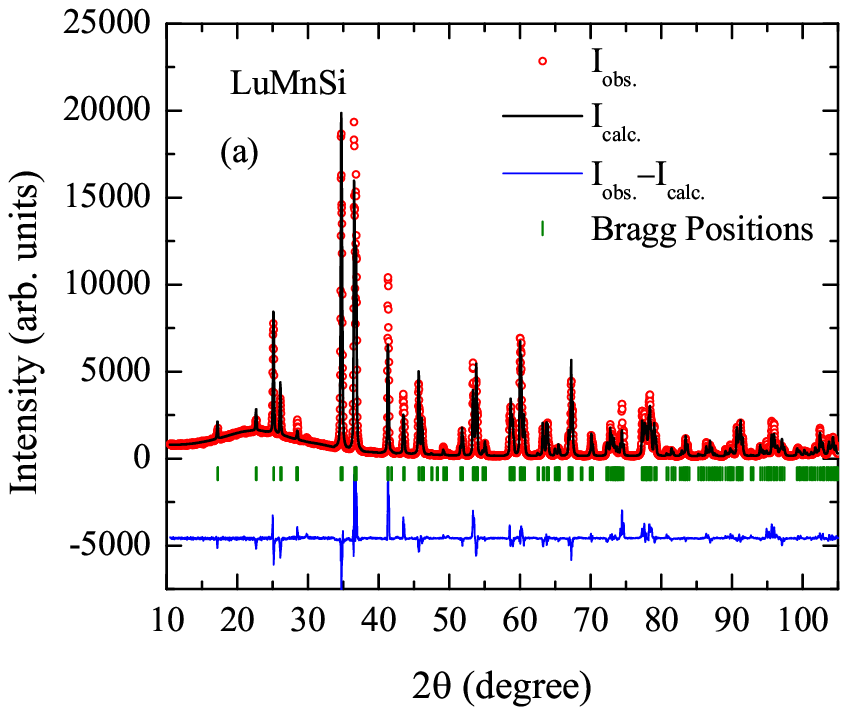}
	\includegraphics[width=2.7in]{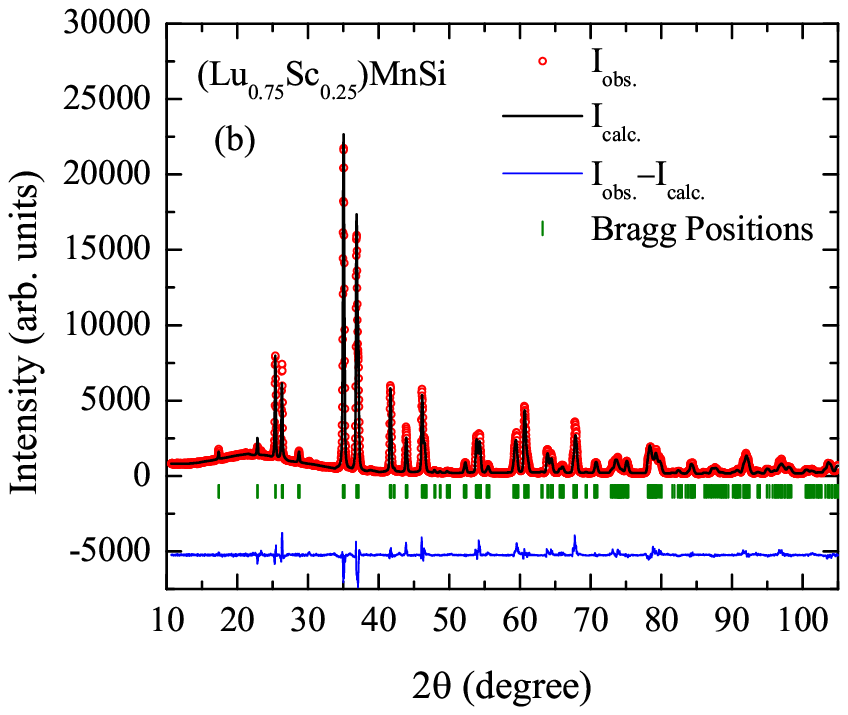}
	\includegraphics[width=2.7in]{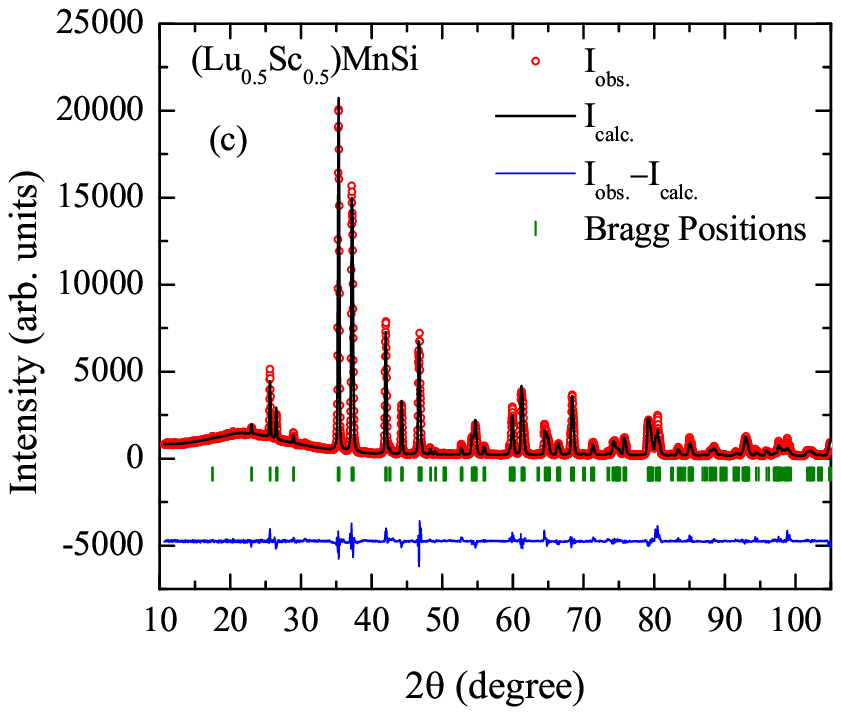}
	\caption{(Color online) Room-temperature powder XRD patterns of (a) LuMnSi, (b) Lu$_{0.75}$Sc$_{0.25}$MnSi and (c) Lu$_{0.50}$Sc$_{0.50}$MnSi.  The data are shown as open red circles, the Rietveld refinement fits as the black lines, the calculated Bragg peak positions as the vertical green tick marks, and the fit devations at the bottom as blue lines.}
	\label{fig:ScDoped_XRD}
\end{figure}

\begin{figure}
	\includegraphics[width=3.3in]{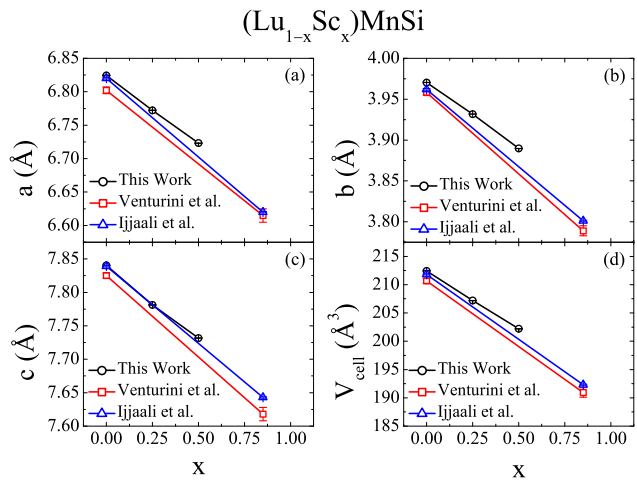}
	\caption{(Color online) Lattice parameters (a) $a$, (b) $b$ and (c) $c$ versus composition $x$ in \lumnsi. Panel (d) shows the unit cell volume $V_{\rm cell}$ versus $x$. Also plotted for comparison are corresponding data from Venturini et al.\cite{Venturini1997} and Ijjaali et al.\cite{Ijjaali1999}  All lines are guides to the eye.}
	\label{fig:crystallographic_comparison}
\end{figure}

The room-temperature powder XRD patterns for \lumnsi\ ($x=0$, 0.25, and 0.50) are shown in Fig.~\ref{fig:ScDoped_XRD}.  The crystal data for the primitive orthorhombic TiNiSi-type structure (space group {\it Pnma}) reported in Refs.~\onlinecite{Venturini1997} and~\onlinecite{Ijjaali1999} were used as the starting point for the Rietveld refinements. Our results confirm this structure for our samples.  The refinements are shown in Fig.~\ref{fig:ScDoped_XRD}. Figure~\ref{fig:crystallographic_comparison} shows the dependence of the lattice parameters on composition. An approximately linear decrease is observed in the $a$, $b$, and $c$ lattice parameters versus Sc concentration~$x$. Table~\ref{table:XRD} contains listings of the lattice parameters and qualities of fit and Table~\ref{table:XRD_coordinates} reports the atomic positions and interatomic distances that may be relevant to the magnetic structure.  Our crystal data are in satisfactory agreement with those reported previously,\cite{Venturini1997, Ijjaali1999} which are also listed in the two tables for comparison.

\begin{table*}
 \caption{\label{table:XRD} Crystallographic properties of the (Lu$_{1-x}$Sc$_{x}$)MnSi system at room temperature (primitive orthorhombic TiNiSi-type structure: space group \emph{Pnma}). Listed for each composition $x$ are the unit cell dimensions $a$, $b$, and $c$; unit cell volume $V_{\rm cell}$; and the Rietveld quality-of-fit parameters $R_{\rm p}$, $R_{\rm wp}$, and $\chi^2$.}
 \begin{ruledtabular}		
		\begin{tabular}{c l l l l l l l  c}
		$x$ 	& $a$ (\AA )	& $b$ (\AA )	& $c$ (\AA )	& $V_{\rm cell}$ (\AA$^3$)	& $R_{\rm p}$	& $R_{\rm wp}$	& $\chi^2$	& Ref. \\ \hline
		0.00	& 6.8246(2)		& 3.9704(2)		& 7.8403(2)		& 212.44(2)					& 11.0			& 23.5			& 19.5  	& This work\\
				& 6.820(1)		& 3.962(1)		& 7.839(1)		& 211.8(1)					& 				&				&			& \onlinecite{Ijjaali1999} \\
				& 6.802(5)		& 3.958(3)		& 7.825(4)		& 210.7(5)					&				&				&			& \onlinecite{Venturini1997} \\
		0.25	& 6.7724(3)		& 3.9318(2)		& 7.7813(3)		& 207.20(3)					& 8.31			& 11.4			& 10.1  	& This work  \\
		0.50	& 6.7232(3)		& 3.8898(2)		& 7.7317(3)		& 202.20(3)					& 6.77			& 9.75			& 7.11  	& This work  \\
		0.90	& 6.620(1)		& 3.801(1)		& 7.643(1)		& 192.29(5)					&				&				&			& \onlinecite{Ijjaali1999} \\
				& 6.615(10)		& 3.789(6)		& 7.618(10)		& 190.9(9)					&				&				&			& \onlinecite{Venturini1997}	\\				
		\end{tabular}
 \end{ruledtabular}
\end{table*}

 \begin{table*}
 \caption{\label{table:XRD_coordinates} Atomic coordinates of the (Lu$_{1-x}$Sc$_{x}$)MnSi system at room temperature. The system crystallizes in space group \emph{Pnma} with $Z=4$ formula units per unit cell.  All atoms occupy $4c$ positions with atomic coordinates Lu: ($x_{\rm Lu/Sc}$, 1/4, $z_{\rm Lu/Sc}$), Mn: ($x_{\rm Mn}$, 1/4, $z_{\rm Mn}$), and Si: ($x_{\rm Si}$, 1/4, $z_{\rm Si}$). Also listed are the interatomic distances (in \AA) that may be relevant to the magnetic ordering of the system.  A listed Mn-Si-Mn distance is the sum of the respective Mn to Si and Si to Mn distances.}
 \begin{ruledtabular}		
		\begin{tabular}{c l l l l l l l l l c}
 		$x$ 	& $x_{\rm Lu/Sc}$ 	& $z_{\rm Lu/Sc}$ 	& $x_{\rm Mn}$ 	& $z_{\rm Mn}$ 	& $x_{\rm Si}$ 	& $z_{\rm Si}$	& $d_{\text{Mn$_2$-Mn$_4$}}$	& $d_{\text{Mn$_1$-Si$_1$-Mn$_2$}}$	& $d_{\text{Mn$_1$-Si$_1$-Mn$_4$}}$ & Ref. \\ \hline
		0.00	& 0.0251(3)			& 0.6803(3)			& 0.1394(8)		& 0.0522(6)		& 0.268(2)		& 0.387(2)		& 2.87(1)	& 5.22(3)	& 5.34(4) 	& This work \\
				& 0.02488(9)		& 0.67835(8)		& 0.1378(3)		& 0.0572(3)		& 0.2742(6)		& 0.3765(6)		& 2.874(3)	& 5.18(2)	& 5.20(2)	& \onlinecite{Ijjaali1999} \\
				& 0.034(3)			& 0.674(3)			& 0.159(8)		& 0.052(5)		& 0.283(6)		& 0.385(5)		& 3.0(1)	& 5.1(2)	& 5.3(3)	& \onlinecite{Venturini1997}\\
		0.25	& 0.0251(2)			& 0.6791(3)			& 0.1335(5)		& 0.0565(5)		& 0.280(1)		& 0.3691(8)		& 2.812(7)	& 5.14(3)	& 5.09(3) 	& This work \\
		0.50	& 0.0258(2)			& 0.6808(3)			& 0.1388(5)		& 0.0627(4)		& 0.2708(9)		& 0.3750(6)		& 2.864(6)	& 5.07(2)	& 5.09(3) 	& This work \\
		0.90	& 0.0281(1)			& 0.6756(1)			& 0.1378(2)		& 0.0560(1)		& 0.2696(3)		& 0.3711(3)		& 2.770(2)	& 5.008(8)	& 5.06(2)	& \onlinecite{Ijjaali1999} \\
				& 0.027(2)			& 0.668(2)			& 0.133(9)		& 0.051(5)		& 0.262(9)		& 0.368(6)		& 2.7(2)	& 5.0(3)	& 5.1(3)  	& \onlinecite{Venturini1997} \\
		\end{tabular}
 \end{ruledtabular}
\end{table*}

\begin{figure}
\includegraphics[width=3in]{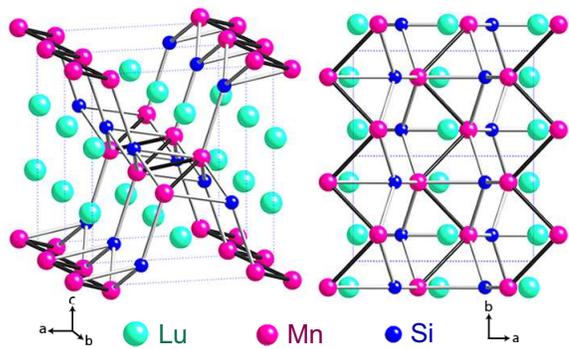}
\caption{(Color online) The structure of LuMnSi viewed from two different perspectives. Two unit cells are shown in each panel with the edges denoted by the thin blue dotted lines. The dark bonds are intrachain Mn-Mn bonds and the lighter ones are interchain Mn-Si-Mn bonds. The Mn zigzag chains are evident in each panel.}
\label{fig:Structure}
\end{figure}

\begin{figure}
\includegraphics[width=2.5in]{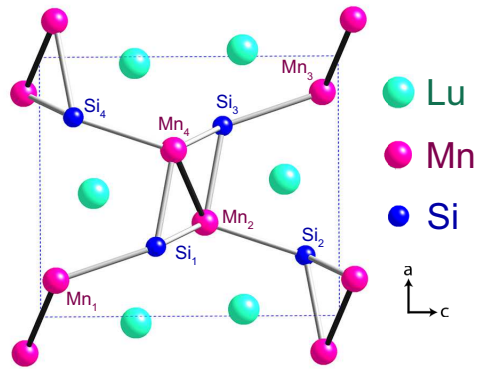}
\caption{(Color online) View of the LuMnSi crystal structure showing the important interactions between the Mn atoms.  The black lines are the direct Mn-Mn interactions and the gray lines are the indirect Mn-Si-Mn interactions.}
\label{fig:Interactions}
\end{figure}

The structure of the \lumnsi\ system is shown in Fig.~\ref{fig:Structure}. The shortest interatomic distances (``bonds'') that are likely most important in determining the magnetic properties are shown, and are more clearly highlighted in Fig.~\ref{fig:Interactions}. These figures show that the Mn atoms are arranged in zigzag chains with the intrachain bonds shown in black. These chains of Mn spins were proposed in Ref.~\onlinecite{Venturini1997} to interact through superexchange interactions via the Si atoms.  These latter interactions are denoted by gray lines in Fig.~\ref{fig:Interactions}. The lengths of these bonds are listed in Table~\ref{table:XRD_coordinates}.  However, if the magnetism arises from local Mn magnetic moments, undoubtedly RKKY interactions between the Mn spins are important in this metallic system.

\section{\label{Transport} Electrical Resistivity}

\begin{figure}
\includegraphics[width=3in]{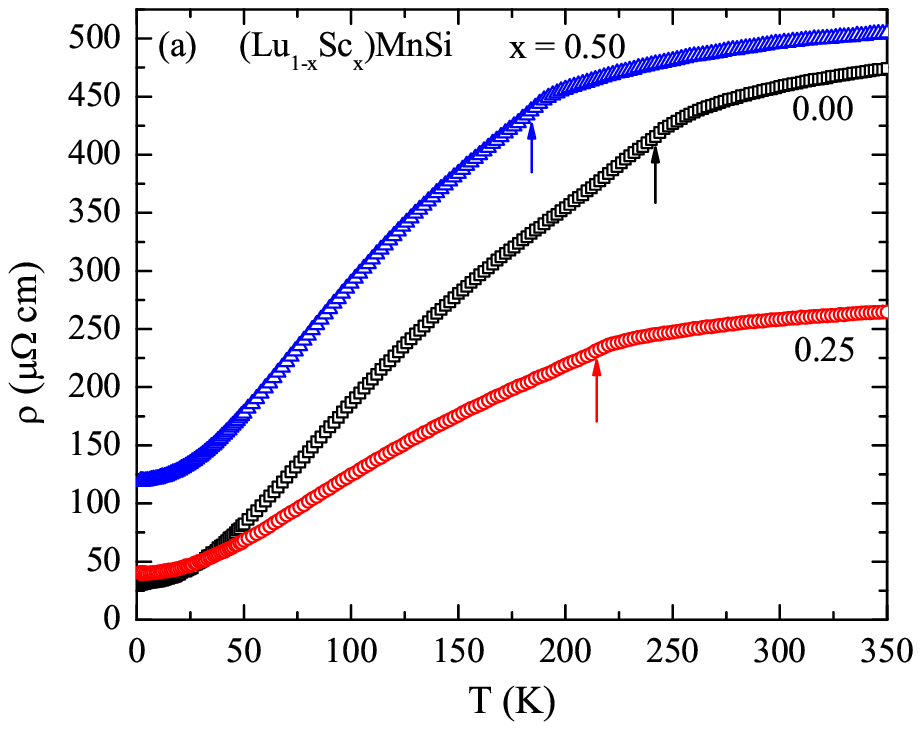}
\includegraphics[width=3in]{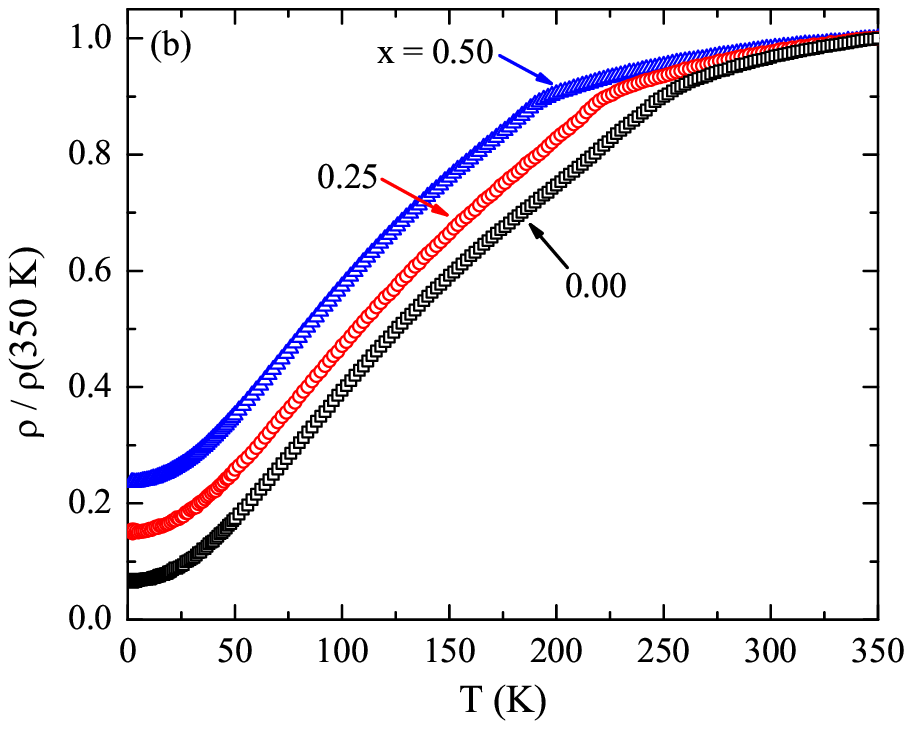}
\includegraphics[width=3in]{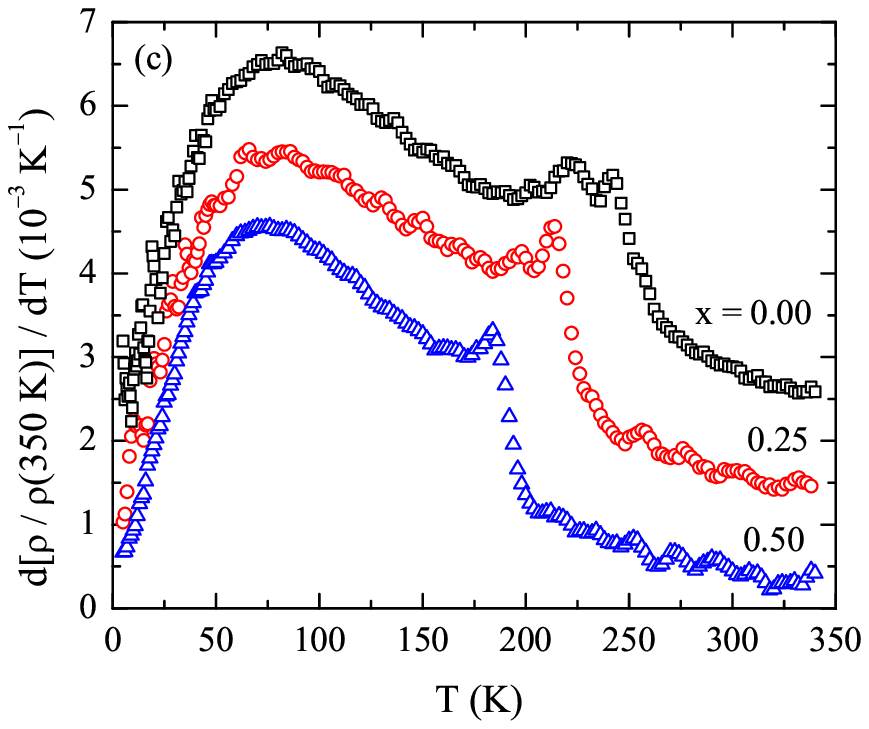}
\caption{(Color online) (a) $\rho$ versus $T$ for the \lumnsi\ system. Arrows indicate $T_{\rm N}$ as determined from the highest $T$ peak of $d\rho/dT$ in (c). (b) $\rho/\rho(350\text{ K})$ versus $T$. (c) Temperature derivative of $\rho/\rho(350\text{ K})$  \{ $d[\rho/\rho(350\text{ K})]/dT$\} to emphasize the change in slope at $T_{\rm N}$. Only the data above $T=5$~K are plotted. For clarity, the data sets $x = 0$ and 0.25 are offset vertically by 2 and $1\times 10^{-3}$~K$^{-1}$, respectively.}
\label{fig:rho}
\end{figure}

The $\rho(T)$ data of our polycrystalline samples of \lumnsi\ ($x=0$, 0.25, 0.50) are presented in Fig.~\ref{fig:rho}(a). The positive slopes and the magnitudes of $\rho(T)$ indicate that the samples are metallic.  Each data set shows a distinct maximum in slope at a temperature $T_{\rm N}$ as marked by the vertical arrows, which we determine below to be the AF transition temperature.  This increase in slope below $T_{\rm N}$ is likely caused by the loss of spin-disorder scattering as ordered moments in the samples increase with decreasing~$T$\@.  To more directly compare the $T$ dependences of $\rho$, the data are normalized by $\rho(300~K)$ and replotted in Fig~\ref{fig:rho}(b).

In order to accurately determine $T_{\rm N}$ from the $\rho(T)$ data, the temperature derivatives of the normalized data in Fig~\ref{fig:rho}(b) are shown in Fig.~\ref{fig:rho}(c), where a sliding least squares fit over a small temperature window was used to calculate the derivative at each~$T$\@. Two peaks are observed in Fig.~\ref{fig:rho}(c) near $T_{\rm N}$ for $x=0$ and~0.25. The lower-$T$ peak may signal a change in the magnetic structure in these two compositions.  The higher-$T$ peak is determined from other measurements below to be $T_{\rm N}$ as indicated by the vertical arrows in Fig~\ref{fig:rho}(a). The values of $T_{\rm N}$ thus determined for compositions $x=0$, 0.25, and 0.50 are 242, 214 and 183~K, respectively.  These values are in reasonable agreement with the values of 246, 215 and 188~K, respectively, that we determined from heat capacity measurements in Sec.~\ref{HC} below.  The data in Fig.~\ref{fig:rho}(c) also show a transition somewhat below $T_{\rm N}$ for $x=0$ and~0.25 that we speculate is due to a spin reorientation transition.  In Fig.~\ref{fig:chi} below we show that the second transition also occurs for $x=0.5$ and that all three are completely suppressed in a field $H=5.5$~T\@.

\section{\label{Magnetic} Magnetization and Magnetic Susceptibility}

\subsection{\label{Eq:MFT} Molecular Field Theory of the Magnetic Susceptibility of a Planar Helical  Antiferromagnetic Structure}

Before presenting the experimental $\chi(T)$ data in the following section, we review the molecular field thoery (MFT) predictions used to analyze the data.\cite{Johnston2012} These predictions are based on a local-moment Heisenberg model where all spins are identical and crystallographically equivalent, as occurs for the Mn spins in the \lumnsi\ system.  The MFT that we use here for a coplanar helix AF structure is formulated in terms of a $J_0$-$J_{z1}$-$J_{z2}$ model shown in Fig.~1 of Ref.~\onlinecite{Johnston2012}, where $J_0$ is the sum of all the interactions of a given spin with spins in the same layer, $J_{z1}$ the sum of the interactions of a given spin to all spins in a nearest-neighbor layer along the helix or cycloid axis, and $J_{z2}$ the sum of the interactions of a given spin to all spins in a next-nearest neighbor layer.\cite{Johnston2012}  Here a ``layer'' is defined as a layer of Mn spins that in the helical AF-ordered state consists of ferromagnetically-aligned Mn ordered moments, where the layer is perpendicular to the helix axis.  The direction of the in-plane ordered moments rotates by a radian angle of $kd$ (see below) from layer to layer along the $z$~axis.

In the paramagnetic state at $T \geq T_{\rm N}$, the measured $\chi(T)$ is a $T$-independent orbital susceptibility $\chi_0$ plus the spin susceptibility given by a Curie-Weiss law, i.e., 
\be
\chi(T) = \chi_0 + \frac{C}{T-\theta_{\rm p}},
\label{CWchi}
\ee
where $\chi_0$ can be anisotropic and generally consists of the diamagnetic core, anisotropic paramagnetic Van Vleck, and conduction carrier Landau diamagnetic susceptibilities.  Also, $\theta_{\rm p}$ is the Weiss temperature, the Curie constant $C$ is
\be
C=\frac{Ng^2S(S+1)\mu_{\rm B}^2}{3k_{\rm B}},
\ee
$N$ is the number of spins, $g$ is the spectroscopic splitting factor ($g$-factor), $\mu_{\rm B}$ is the Bohr magneton and $k_{\rm B}$ is Boltzmann's constant.  For the helical AF structure, using MFT one can write $\theta_{\rm p}$ and $T_{\rm N}$ in terms of the above-defined $J_0$, $J_{z1}$, $J_{z2}$ interactions as\cite{Johnston2012}
\bse
\label{Eqs:helix}
\bea
k_{\rm B} \theta_{\rm p} &=& -\frac{S(S+1)}{3}(J_0+2J_{z1}+2 J_{z2}) ,
\label{eq:exchange_thetaP}\label{Eq:thetap}\\
k_{\rm B} T_{\rm N} &=& -\frac{S(S+1)}{3} \label{Eq:TN} \\
&& \times [J_0+2J_{z1} \cos(kd)+2 J_{z2} \cos(2kd)],\nonumber
\eea
where $k$ is the magnitude of the helix wavevector, $d$ is the distance between adjacent ferromagnetically-aligned layers of spins, and therefore $kd$ is the turn angle in radians along the $z$-axis between spins in adjacent spin layers.  After writing the classical energy of the helix at $T=0$ in terms of the above exchange constants and $kd$, minimizing the energy with respect to $kd$ yields the relationship 
\be
\cos(kd) = -\frac{J_{z1}}{4 J_{z2}}.
\label{Eq:kd}
\ee
\ese
In general, $\theta_{\rm p}$ and $ T_{\rm N}$ in Eqs.~(\ref{Eq:thetap}) and~(\ref{Eq:TN}), respectively, are known from the experimental $\chi(T)$ data and $S$ can be estimated from these data or other considerations.  The value of $kd$ in Eq.~(\ref{Eq:kd}) can be obtained from the experimental value of $\bar{\chi}_{xy}(T=0)$ using Eq.~(\ref{eq:chi_xyT0}) below.  Then the values of $J_0,\ J_{z1},\ J_{z2}$ are obtained by solving the three simultaneous Eqs.~(\ref{Eqs:helix}) using the known values of $\theta_{\rm p}$, $ T_{\rm N}$, $S$ and $kd$.

When $H$ is applied perpendicular to the ordering plane of any collinear or planar noncollinear AFM structure containing identical crystallographically equivalent spins interacting by Heisenberg exchange, MFT predicts that the perpendicular magnetic susceptibility $\chi_\perp$ at $T \leq T_{\rm N}$ is constant and equal to the value at $T_{\rm N}$,\cite{Johnston2012} i.e., the dimensionless reduced perpendicular susceptibility $\bar{\chi}_\perp(T)$ is
\be
\bar{\chi}_\perp(T\leq T_{\rm N}) \equiv \frac{\chi_\perp(T\leq T_{\rm N})}{\chi(T_{\rm N})} = 1.
\label{eq:chi_perp}
\ee
When $H$ is applied in the plane of the spins of a planar ($xy$) noncollinear AF structure, MFT predicts that the dimensionless reduced susceptibility is\cite{Johnston2012} 
\bse
\be
\bar{\chi}_{xy}(t) \equiv \frac{\chi_{xy}(T)}{\chi(T_{\rm N})} = \frac{(1+\tau^*+2f+4B^*)(1-f)}{2[(\tau^*+B^*)(1+B^*)-(f+B^*)^2]},
\label{eq:chi_xy}
\ee
where the dimensionless reduced variables are
\begin{align}
\tau^* &= \frac{(S+1)t}{3 B_S'(y_0)}, \qquad y_0 = \frac{3 \bar{\mu}_0}{(S+1)t},\\
B^* &= 2(1-f)\cos(kd)[1+\cos(kd)]-f ,\label{Eq:B*}\\
f &= \frac{\theta_{\rm p}}{T_{\rm N}},\label{Eq:f}\\
t &= \frac{T}{T_{\rm N}}.
\end{align}
The reduced ordered moment in appled field $H=0$ is defined as $\bar{\mu}_0(t) \equiv \mu_0(t)/\mu_0(0)$, where $\mu_0(t)$ is the temperature-dependent ordered moment below~$T_{\rm N}$\@.  The value of $\bar{\mu}_0$ is obtained by numerically solving
\be
\bar{\mu}_0 = B_S(y_0),
\ee
and the value of $f$ is usually uniquely defined from the measured values of $\theta_{\rm p}$ and $T_{\rm N}$.  The sign of $f$ is the same as the sign of $\theta_{\rm p}$ and can therefore be either positive or negative.

The expression for $B^\ast$ in Eq.~(\ref{Eq:B*}) applies specifically to a planar helix with the helix axis being the $z$~axis which is perpendicular to the $xy$~plane in which the ordered moments are aligned for $T < T_{\rm N}$.  At $T=0$, Eq.~(\ref{eq:chi_xy}) yields
\be
\bar{\chi}_{xy}(T=0)  =  \frac{1}{2[1+2\cos (kd) + 2\cos^2(kd)]},
\label{eq:chi_xyT0}
\ee
which does not contain the parameter $f$.  However, the temperature dependence of $\bar{\chi}_{xy}$ does depend on~$f$ according to Eq.~(\ref{eq:chi_xy}). Our unconventional definition of the Brillouin function and its derivative are
\bea
B_S(y_0) &=& \frac{1}{2S}\bigg\{(2S+1) {\rm coth}\left[ (2S+1) \frac{y_0}{2}\right]\nonumber\\*
&&\hspace{1.6in} -\ {\rm coth}\left[\frac{y_0}{2}\right]\bigg\}, \nonumber\\
B_S'(y_0) &= & \frac{1}{4S}\bigg\{ {\rm csch}^2 \left(\frac{y_0}{2}\right)\\
&&\hspace{0.5in}  -\ (2S+1)^2 {\rm csch}^2 \left[(2S+1)\frac{y_0}{2}\right] \bigg\}. \nonumber
\eea
\ese

\begin{figure}
\includegraphics[width=3.3in]{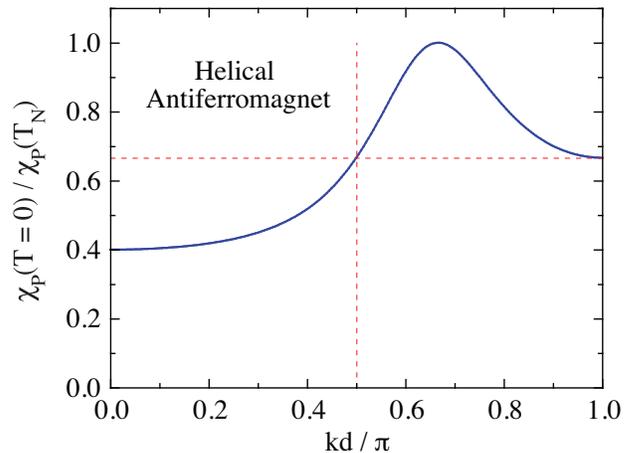}
\caption{(Color online) Reduced powder susceptibility $\bar{\chi}_{\rm P}(T=0)\equiv \chi_{\rm P}(T=0)/\chi_{\rm P}(T_{\rm N}$) versus the turn angle $kd$ in radians for a helical Heisenberg antiferromagnet according to Eq.~(\ref{eq:chi_p_t=0}) (solid blue curve).  For the upper right quadrant with $\pi/2 < kd < \pi$ defined by the red dashed horizontal and vertical lines, there are two possible values of  $kd$ for a given value of $\bar{\chi}_{\rm P}(T=0)$.  The special point at the maximum of the curve with $kd=2\pi/3$ and $\bar{\chi}_{\rm P}(T=0)=1$ corresponds to 120\degr\ ordering of a triangular lattice antiferromagnet for which the single-crystal $\chi$ is isotropic and independent of~$T$ and spin~$S$ below~$T_{\rm N}$,\cite{Johnston2012} and hence the powder average is also independent of~$T$ and~$S$ below~$T_{\rm N}$.}
\label{Fig:Helix_chi_vs_kd_powder}
\end{figure}

\begin{figure}
	\includegraphics[width=2.9in]{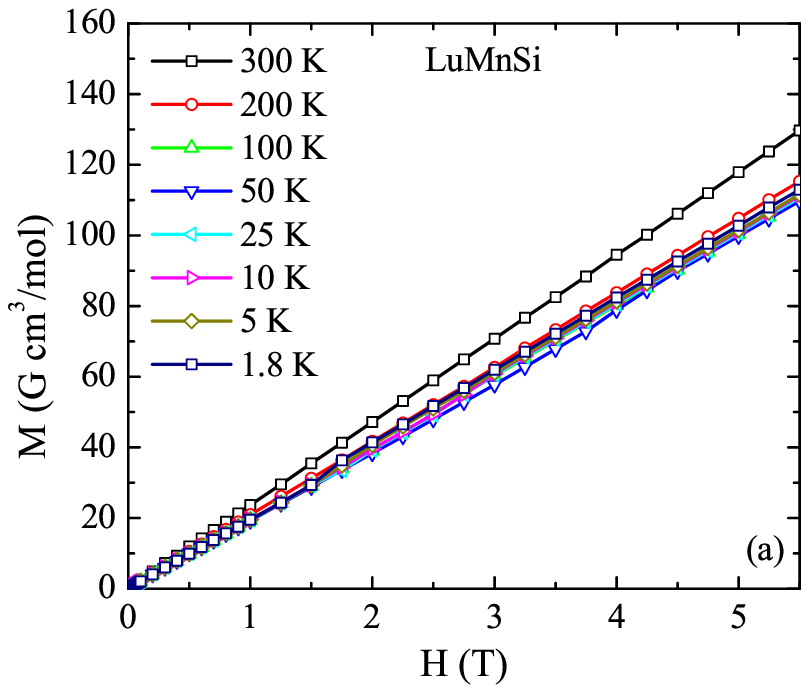}
	\includegraphics[width=2.9in]{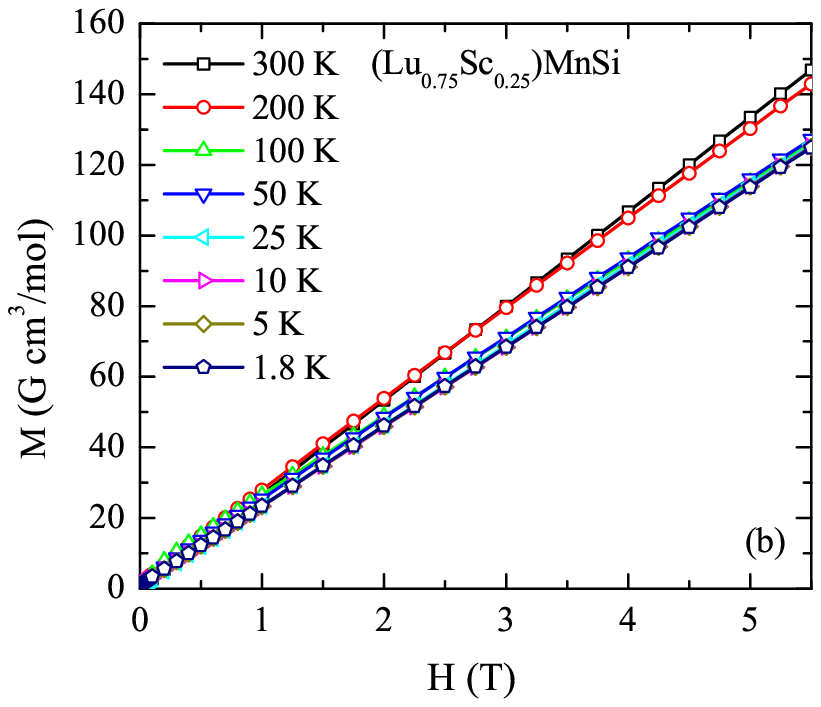}
	\includegraphics[width=2.9in]{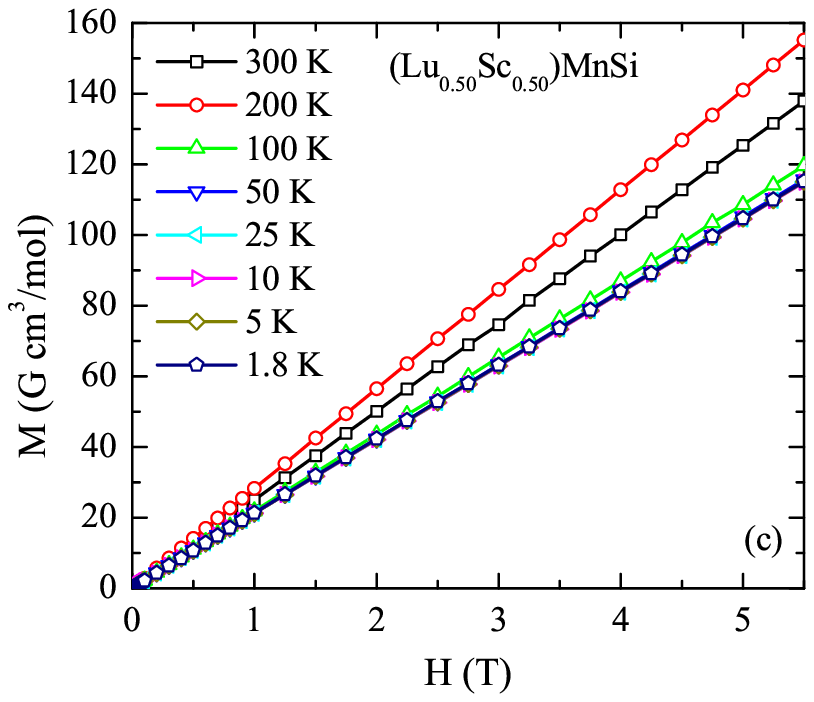}
	\caption{(Color online) Magnetization $M$ versus applied magnetic field $H$ isotherms for samples in the \lumnsi\ system for compositions (a) $x=0.00$ (b) $x=0.25$, and (c) $x=0.50$. The lines connecting the points are guides to the eye.}
	\label{fig:MH}
\end{figure}

For a polycrystalline sample, one measures the powder-averaged susceptibility $\chi_{\rm p}$ given in reduced units using Eq.~(\ref{eq:chi_perp}) as
\bse
\be
\bar{\chi}_{\rm P}(T) \equiv \frac{{\chi}_{\rm P}(T)}{\chi(T_{\rm N})}= \frac{1}{3} [\bar{\chi}_\perp + 2 \bar{\chi}_{xy}(T)] = \frac{1}{3} [1 + 2 \bar{\chi}_{xy}(T)] .
\label{eq:chi_p}
\ee
Then using Eqs.~(\ref{eq:chi_xyT0}) and~(\ref{eq:chi_p}), in the ordered helical AF state at $T=0$ one obtains
\be
\bar{\chi}_{\rm P}(T=0) = \frac{1}{3} \left[ 1 +  \frac{1}{1+2\cos (kd) + 2\cos^2(kd)} \right],
\label{eq:chi_p_t=0}
\ee
\ese
which allows the helix turn angle $kd$ to be determined from the measured values of $\chi_{\rm P}(T=0)$ and $\chi(T_{\rm N})$.  A plot of $\bar{\chi}_{\rm P}(T=0)$ versus $kd$ according to Eq.~(\ref{eq:chi_p_t=0}) is shown in Fig.~\ref{Fig:Helix_chi_vs_kd_powder}.  A turn angle~$kd$ less than $\pi/2$~rad indicates that the dominant interactions in the system are ferromagnetic, since in this case there is a component of the ordered moments in adjacent layers that are in the same direction.  Conversely, a turn angle~$kd$ greater than $\pi/2$~rad indicates that the dominant interactions in the system are antiferromagnetic.  

Figure~\ref{Fig:Helix_chi_vs_kd_powder} shows that for $2/3 < \bar{\chi}_{\rm P}(T=0) < 1$ as found below for the \lumnsi\ system, the derived value of $kd$ is not unique since for a $\bar{\chi}_{\rm P}(T=0)$ value in this range Eq.~(\ref{eq:chi_p_t=0}) yields two possible values of $kd$ with $\pi/2 < kd < \pi$.  A corresponding plot of $\bar{\chi}_{xy}(T=0)$ versus $kd$ from Eq.~(\ref{eq:chi_xyT0}) that can be used to analyze single-crystal $\bar{\chi}_{xy}(T=0)$ data is shown in Fig.~2(a) of Ref.~\onlinecite{Johnston2012}.

\subsection{Magnetization and Magnetic Susceptibility Measurements and Analysis}

\begin{figure}
	\includegraphics[width=3in]{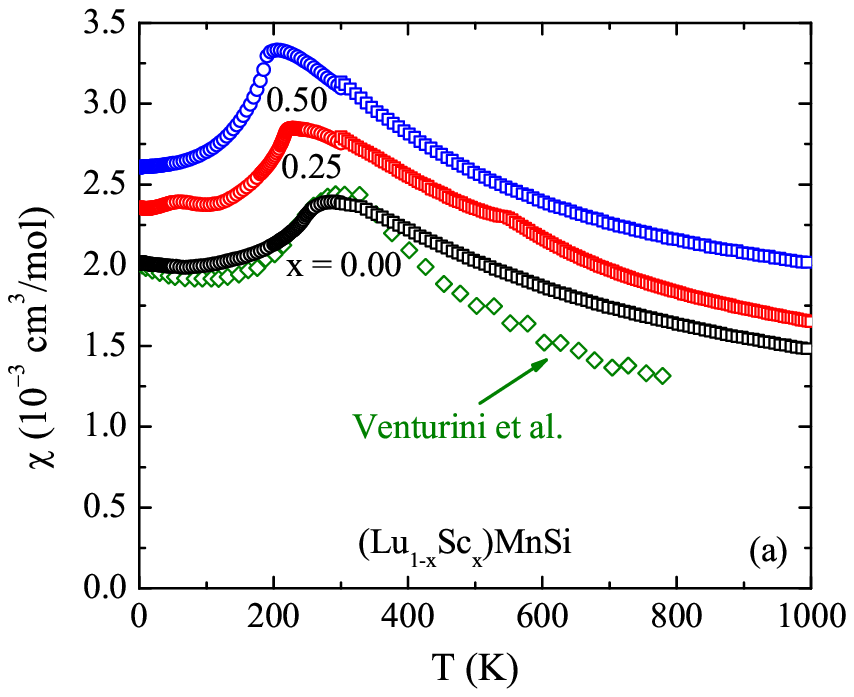}
	\includegraphics[width=3in]{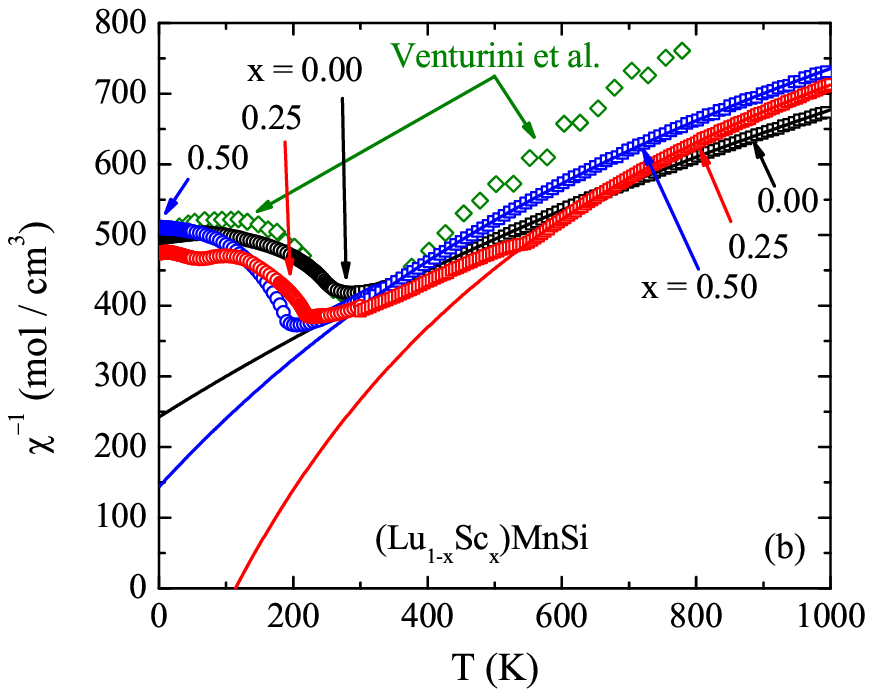}
	\includegraphics[width=3in]{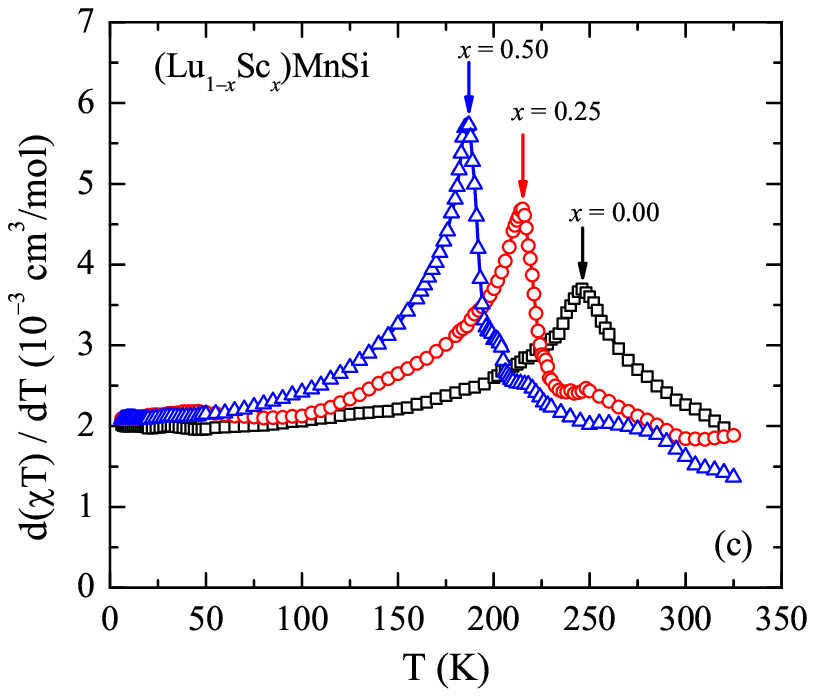}
	\caption{(Color online) (a,b) Magnetic susceptibility $\chi$ versus temperature $T$ for samples in the \lumnsi\ system. Data below 300~K were measured on a SQUID magnetometer (MPMS, circles) and data above 300~K were measured on a VSM (squares). The SQUID data were measured in an applied field $H=5.5$~T and the VSM in $H=3.0$~T\@. For clarity, the data for $x=0.25$ and $x=0.50$ are offset by $0.25\e{-3}$ and $0.65\e{-3}$~cm$^3$/mol, respectively.  One data point is plotted for every five measured for the composition $x = 0.25$ and one of every twenty for $x = 0.00$ and 0.50. The solid curves are fits by the modified Curie-Weiss law in Eq.~(\ref{CWchi}).  For comparison, also shown are the data from Ref.~\onlinecite{Venturini1997} for $x=0$ in $H=1.2$~T\@. (c) Plot of $d(\chi T)/dT$ versus $T$\@. The numerical derivative was evaluated using a sliding window least-squares fit.}
	\label{fig:high_temp_chi}
\end{figure}

\begin{figure}
	\includegraphics[width=3in]{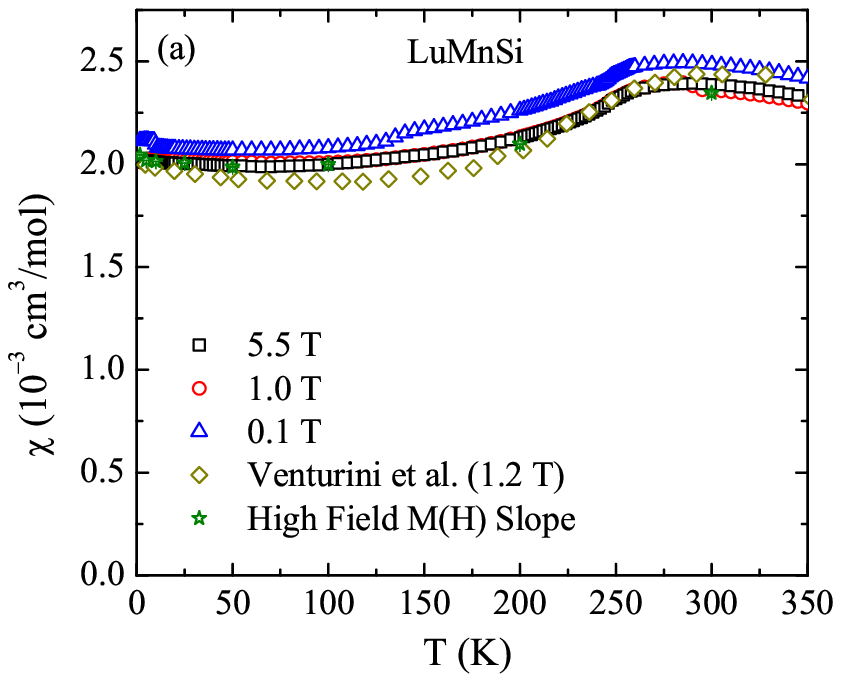}
	\includegraphics[width=3in]{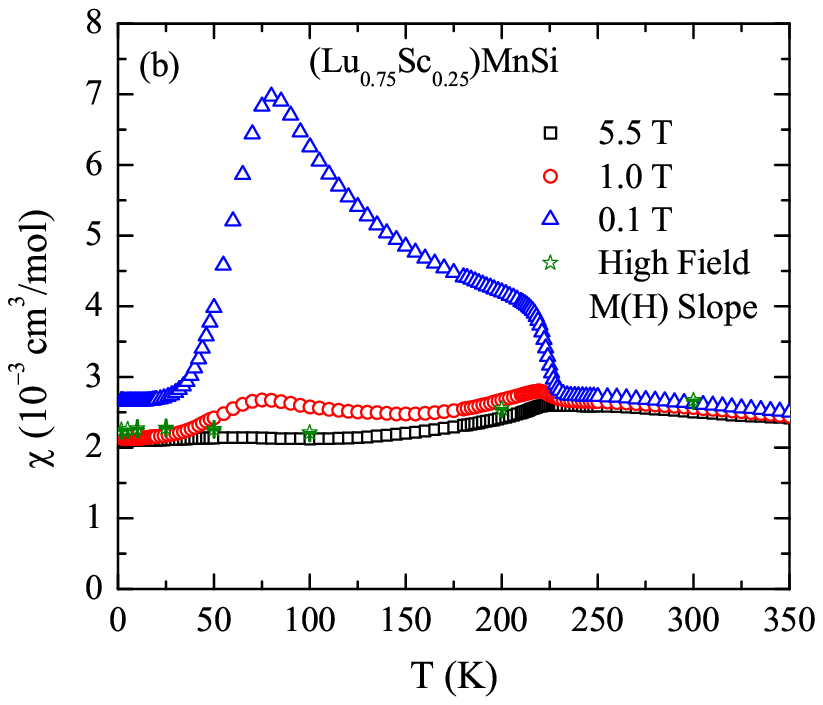}
	\includegraphics[width=3in]{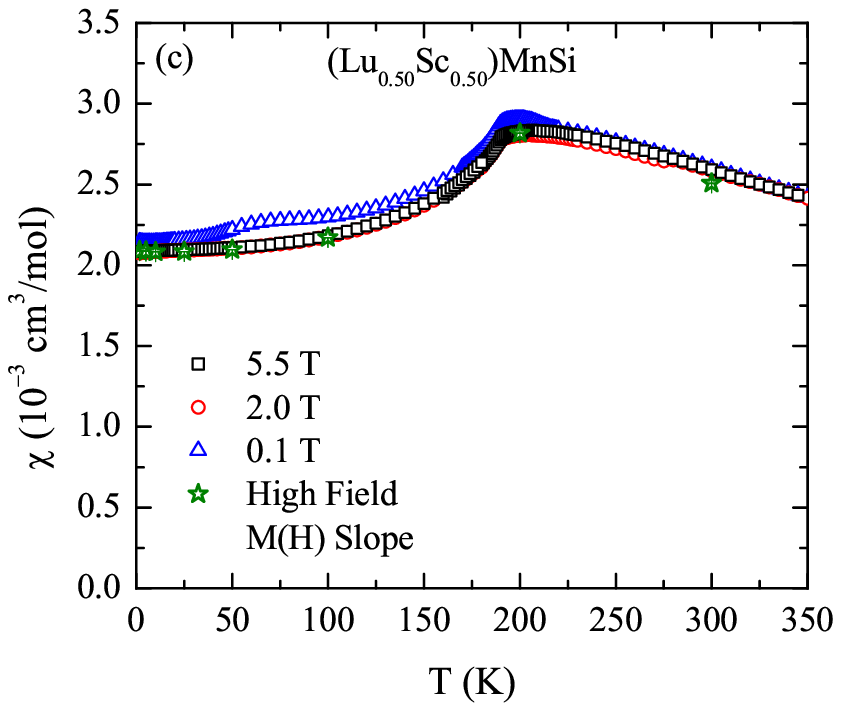}
	\caption{(Color online) Magnetic susceptibility $\chi\equiv M/H$ versus temperature $T$ data for samples in the \lumnsi\ system with compositions (a) $x=0.00$, (b) $x=0.25$, and (c) $x=0.50$. In (a) data from Ref.~(\onlinecite{Venturini1997}) are also plotted for comparison. The values of the high-field $M(H)$ slopes were obtained from a linear fit of the $M(H)$ data in Fig.~\ref{fig:MH} above $H=1$~T\@. The low-$T$ transition for each sample (possibly a spin-reorientation transition) is seen to be suppressed by $H = 5.5$~T\@.}
	\label{fig:chi}
\end{figure}

\begin{figure}
	\includegraphics[width=3in]{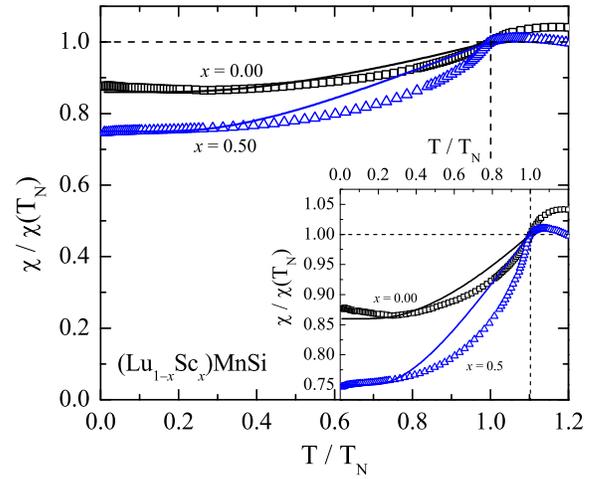}
	\caption{(Color online) Normalized magnetic susceptibility $\chi\, / \, \chi(T_{\rm N})$ versus normalized temperature $T/T_{\rm N}$ for samples in the \lumnsi\ system.  Solid curves are fits by MFT assuming $S=1$ as described in the text with parameters listed in Table~\ref{table:chi_fits}.}
	\label{fig:chi_comparison}
\end{figure}

\begin{table*}
\caption{\label{table:chi_fits} Parameters obtained by fitting the high-temperature $\chi^{-1}$ versus $T$ data and low-temperature ($T < T_{\rm N}$) $\chi/\chi(T_{\rm N})$ versus $T/T_{\rm N}$ data in Fig.~\ref{fig:chi_comparison}. Listed are: the $T$ range used to fit the high-temperature $\chi^{-1}$ data in Fig.~\ref{fig:high_temp_chi}(b), $T_{\rm N}$ as determined from the peak in $d(\chi T)/dT$, Weiss temperature $\theta_{\rm p}$, the calculated ratio $f \equiv \theta_{\rm p}/T_{\rm N}$, Curie constant $C$ per mole of formula units, Curie constant $C_{\rm Mn}$ per mole of Mn atoms and  helix turn angle $kd$ for the different compositions $x$. Two values of $kd$ are listed for reasons described in the text. Also included are literature data from Ref.~\onlinecite{Venturini1997}. }
\begin{ruledtabular}
\begin{tabular}{cccccccccc}
$x$		& $T$ range  &  $T_{\rm N}$  & $\theta_{\rm p}$ & $f$  & $C$ & $C_{\rm Mn}$   & $\chi_0$  & $kd$  	 	& Ref. \\ 
	& (K)	& (K)	&  (K)	& 	& ($\rm{\frac{cm^3\,K}{mol}}$)& ($\rm{\frac{cm^3\,K}{mol\,Mn}}$)	& ($\rm{\frac{10^{-3}\,cm^3}{mol}}$)	& (deg)	  \\	 \hline
0.00	& 350--1000		& 246			& $-$352(2)				& $-$1.43		& 1.262(5)&	1.262(5)  	& 0.544(3)	& 104 or 139		& This work \\
0.00	&				& 255			& $-$201				& 				& 1.25	&	1.25		& 		&        			& \onlinecite{Venturini1997} \\
0.25	& 550--1000		& 215			& 114(2)				& 0.531			& 0.551(3)&	0.735(4)	& 0.781(2)	& 	98.9 or 147	& This work \\
0.50 	& 300--1000		& 189			& $-$125.2(8)			& $-$0.662		& 0.789(2)& 1.578(4)	& 0.659(2)& 96.5 or 153		& This work \\
0.85	&				& 125			& 166					& 				& 0.374&				& 	& 			 		& \onlinecite{Venturini1997}
\end{tabular}
\end{ruledtabular}
\end{table*}

$M(H)$ isotherms were measured for each of the three  \lumnsi\ ($x=0$, 0.25, 0.50) samples in magnetic fields up to $H=5.5$~T at eight temperatures between 1.8 and 300~K as shown in Fig.~\ref{fig:MH}.  The data show that $M$ is proportional to~$H$ over the entire field range for each of the samples, thus indicating that ferromagnetic or saturable paramagnetic impurities are not present in this temperature range for any of the samples.

The $\chi(T)\equiv M(T)/H$ measurements for 1.8 to 1000~K carried out at various $H$ are shown in Fig.~\ref{fig:high_temp_chi}(a) and the data below 350~K are shown separately in Fig.~\ref{fig:chi}. The inverse susceptibility $\chi^{-1}$ is plotted versus $T$ in Fig.~\ref{fig:high_temp_chi}(b).  
In Figs.~\ref{fig:high_temp_chi}(a), \ref{fig:high_temp_chi}(b) and \ref{fig:chi}(a), our data for the $x=0$ sample are compared with the results of a previous study for the same composition.\cite{Venturini1997} For $T< T_{\rm N}$, good agreement is observed between these data sets.  However the data in the paramagnetic region at $T > T_{\rm N}$ differ significantly.  This resulted in different values for $C$ and $\theta_{\rm p}$ from Curie-Weiss fits to the two respective data sets. 

The $\chi^{-1}(T)$ data in Fig.~\ref{fig:high_temp_chi}(b) were fitted by the inverse of the expression for the susceptibility in Eq.~(\ref{CWchi}).  In the sample with composition $x=0.25$, a break in slope is observed at $T=550$~K\@. The cause of this feature is unknown and Eq.~(\ref{CWchi}) was fitted above this temperature.  The $T$ range of the fits and the parameters obtained are given in Table~\ref{table:chi_fits}.  The parameter $f=\theta_{\rm p}/ T_{\rm N}$ in Eq.~(\ref{Eq:f}) is calculated from the resulting value of $\theta_{\rm p}$ and the $T_{\rm N}$ determined from heat capacity measurements below.

The values of the Curie constant per mole of Mn, $C_{\rm Mn}$, for the $x = 0$ and~0.5 samples are 1.26 to 1.58~${\rm cm^3\,K/mol\,Mn}$, respectively.  Curie constants $C_{\rm Mn} = 1$ and $1.875~{\rm cm^3\,K/mol\,Mn}$ correspond to  Mn spins $S=1$ and~3/2 with $g=2$ in a local moment picture, respectively, indicating that within this picture the Mn spins are rather small.  In the analysis below of the data, we assume that $S=1$ and that $g$ is somewhat greater than 2.  The variability in $C_{\rm Mn}$ with $x$ suggests that \lumnsi\ may be an itinerant spin system. However, itinerant spin systems are often modeled using a local moment picture as carried out for the iron-arsenide high-$T_{\rm c}$ superconductors.\cite{Johnston2010}

The slopes of the proportional $M(H)$ data in Fig.~\ref{fig:MH} were determined by fitting the high-field data with $H\geq 1$~T and are plotted as the filled stars in Fig.~\ref{fig:chi}.  We expected and found that these data are in good agreement with the respective $\chi\equiv M/H$ data for $H > 1$~T\@.  The data in Fig.~\ref{fig:chi} for all three samples exhibit long-range AF order at N\'eel temperatures $T_{\rm N} \sim 200$--250~K\@.  The $T_{\rm N}$ of each sample is the temperature of the maximum of the respective derivative $d(\chi T)/dT$ (Ref.~\onlinecite{Fisher1962}) as plotted in Fig.~\ref{fig:high_temp_chi}(c). The $T_{\rm N}$ for each sample is reported in Table~\ref{table:chi_fits}.

In addition to the AF transition seen in $\chi(T)$ at $T_{\rm N}$, each sample shows an anomaly at a lower~$T$ and at low fields that may reflect a change in the magnetic structure from an incommensurate helical structure (see below) to a commensurate one.  This anomaly is at $\approx 140$, 80 and 70~K for $x=0$, $x=0.25$ and $x=0.50$, respectively.  Neutron diffraction measurements specifically directed at this question are needed to resolve it.  As seen in Fig.~\ref{fig:chi}, the transitions at $T_{\rm N}$ are not significantly affected by fields up to 5.5~T, whereas the lower-$T$ transitions are completely  suppressed by $H=5.5$~T\@.

For the purpose of analyzing $\chi(T=0)$ and the $T$ dependence of $\chi$ for $0<T<T_{\rm N}$, in Fig.~\ref{fig:chi_comparison} are plotted the normalized susceptibilities $\bar{\chi}(T)\equiv \chi(T)/\chi(T_{\rm N})$ of the two samples with $x=0$ and~0.50 in $H=5.5$~T from Fig.~\ref{fig:chi}.  We omit the data for the sample with $x=0.25$ in this figure because of the $\chi(T)$ anomaly at $T=550$~K that drastically affects the value of $f$ for this sample that is needed to analyze the temperature dependence of $\chi$ below~$T_{\rm N}$.

\begin{figure}
	\includegraphics[width=3in]{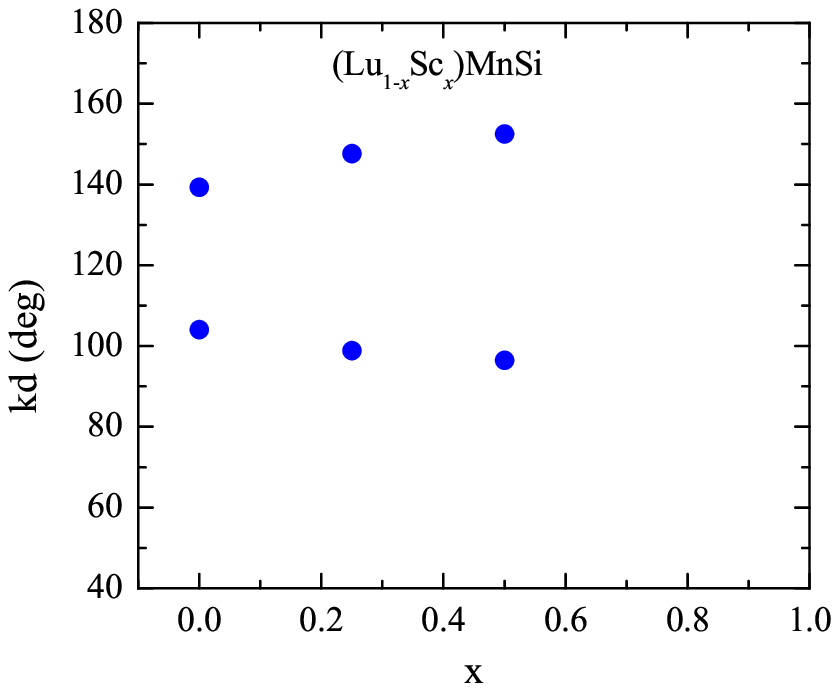}
	\caption{(Color online) Helix turn angle $kd$ versus composition $x$ for samples in the \lumnsi\ system. There are two solutions of Eq.~(\ref{eq:chi_p_t=0}) for $kd$ for each composition as described in the text.}
	\label{fig:kd}
\end{figure}

Within MFT, the powder-averaged and normalized $\bar{\chi}(T=0)$ for a collinear local-moment Heisenberg AF is $\bar{\chi}(T=0)=2/3$, whereas Fig.~\ref{fig:chi} shows that the $\bar{\chi}(T=0)$ values for the three samples are in the range 0.74--0.88.  This difference suggests a noncollinear ground state for each sample at $H=5.5$~T as discussed above.  We therefore assume that the magnetic structure of the ground state is a planar noncollinear helix as proposed in Ref.~\onlinecite{Venturini1997}. As discussed in Sec.~\ref{Eq:MFT}, the turn angle $kd$ between successive layers of spins can be estimated from the normalized value of the powder susceptibility $\bar{\chi}_{\rm P}(T=0)$ according to Eq.~(\ref{eq:chi_p_t=0}) and Fig.~\ref{Fig:Helix_chi_vs_kd_powder}.  From Fig.~\ref{Fig:Helix_chi_vs_kd_powder}, the observed values $\bar{\chi}_{\rm P}(T=0) = 0.72$--0.88 are in the range in which two solutions for $kd$ are possible.  The obtained values of $kd$ are presented in Table~\ref{table:chi_fits} and plotted as a function of $x$ in Fig.~\ref{fig:kd}. 

Taking the smaller of the two possible values of $kd$ (Ref.~\onlinecite{Venturini1997}) for the $x = 0$ and~0.5 samples and the measured values of $f$ for the two samples, the temperature dependences of $\chi_{\rm p}$ in Eq.~(\ref{eq:chi_p}) are fully determined. These MFT predictions assume that $S=1$ as estimated from the above Curie-Weiss fits.  The fits are compared with our observed data in Fig.~\ref{fig:chi_comparison} for $x = 0$ and $x = 0.5$.

The temperature dependences of our data in Fig.~\ref{fig:chi_comparison} are not  described very well by the MFT\@. It was observed in Ref.~\onlinecite{Johnston2012} that fits of such experimental data by the MFT prediction  deviate from the data as the spin quantum number decreases, and this deviation is in the same direction as seen in Fig.~\ref{fig:chi_comparison}.  This occurs because molecular field theory does not include the influence of quantum fluctuations arising from finite spin. Therefore, within the local-moment model the discrepancy between theory and experiment in Fig.~\ref{fig:chi_comparison} is due to the small spin $S\sim 1$ of the Mn atoms. 

The exchange constants $J_0$, $J_{z1}$ and $J_{z2}$ within the helical model for the $x=0$ and~0.5 samples are obtained by solving the system of equations~(\ref{Eqs:helix}), where the input parameters are $S$, $\theta_{\rm p}$, $T_{\rm N}$ and~$kd$. The results are presented in Table~\ref{table:exchange_constants} for both of the possible values of $kd$ for the sample with $x=0$ and~0.5 in Table~\ref{table:exchange_constants}, where we assume $S=1$.  For each composition, the sum of the exchange interactions of a Mn spin with all other spins is $J_0 + J_{z1} + J_{z2} > 0$, comfirming that the dominant interactions in the system are antiferromagnetic as discussed above.

\begin{table}
\caption{\label{table:exchange_constants} Estimates of the exchange constants in the \lumnsi\ system using the assumed $J_0$-$J_{z1}$-$J_{z2}$ model. These are obtained by solving the system of equations in Eqs.~(\ref{Eqs:helix}). Two solutions for $kd$ in Eqs.~(\ref{Eqs:helix}) are given for each composition for a given value of $\bar{\chi}_{\rm P}(T=0)$ as discussed in the text, and hence two sets of exchange constants for each composition. Values for $x=0.25$ are not listed due to an ambiguity caused by the different behavior of $\chi(T)$ for $T > 300$~K\@.}
\begin{ruledtabular}
\begin{tabular}{ccccc}
$x$ & $kd$  & $J_0/k_{\rm B}$  & $J_{z1}/k_{\rm B}$& $J_{z2}/k_{\rm B}$  \\ 
	& (deg)&	(K)			&   (K) 			&  (K)\\\hline
0.00	& 104	& $-$44(1)	& 140.6(5)	& 145.4(5)		\\
		& 139	& $-$57(2)	& 219.8(7)	& 72.8(3)		\\
0.50	& 96.5	& $-$87.3(5)& 42.9(2)	& 94.7(3)		\\
		& 153	& $-$111.7(5)& 116.9(3)	& 32.8(2)			
\end{tabular}
\end{ruledtabular}
\end{table}

\newpage

\section{\label{HC} Heat Capacity}
	 
\begin{figure}
\includegraphics[width=3in]{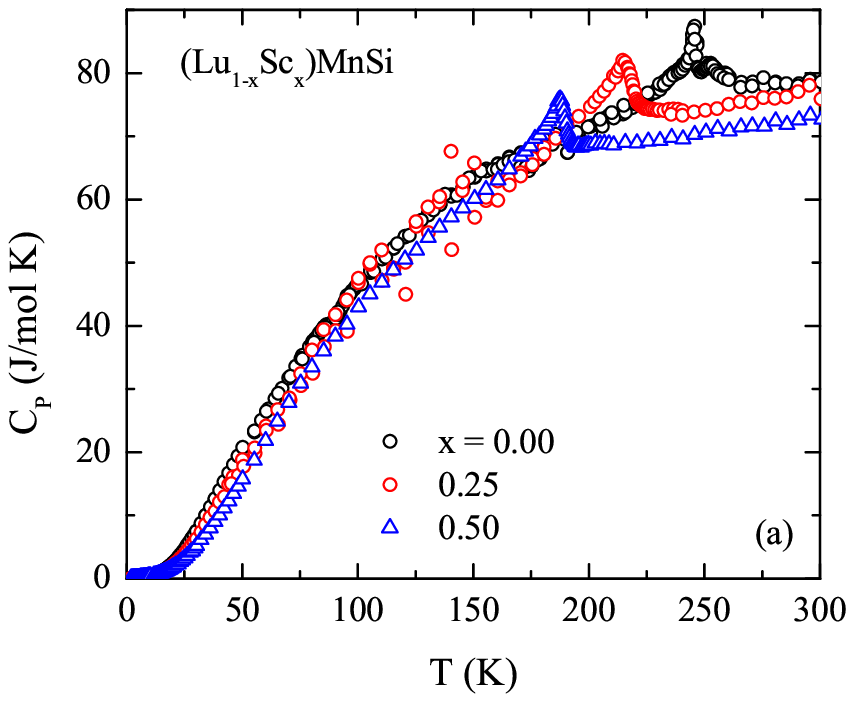}\vspace{0.1in}
\includegraphics[width=3.1in]{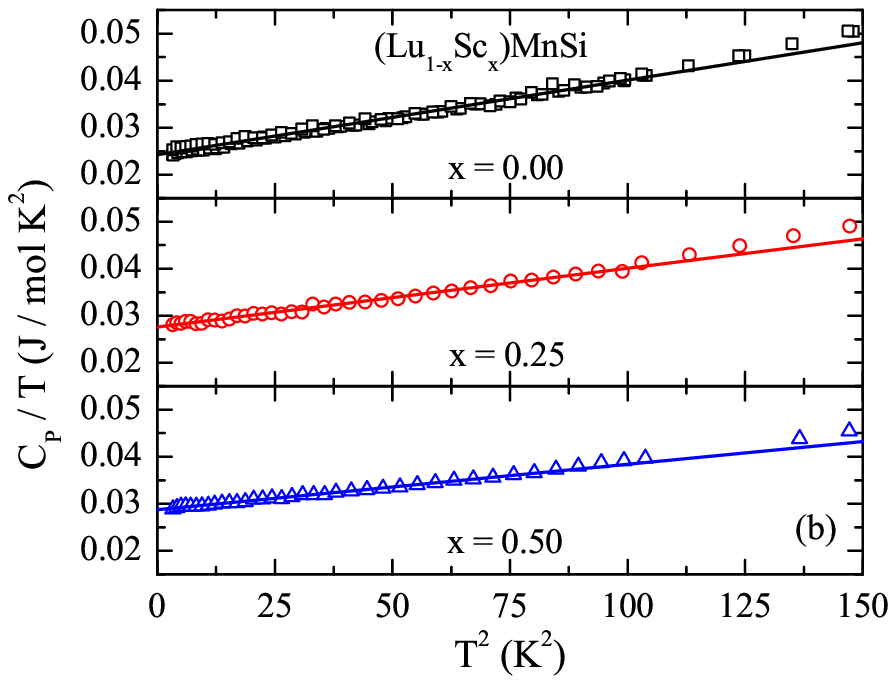}
\caption{(Color online) (a) Heat capacity measured at constant pressure $C_{\rm p}$ versus temperature $T$ for samples in the \lumnsi\ system. (b) Low-$T$ plot of $C_{\rm p}/T$ versus $T^2$ according to Eq.~(\ref{Eq:C/TvsT2Fit}). The straight lines of corresponding color are linear fits and the fit parameters $\gamma$ and $\beta$ are listed in Table~\ref{table:cp_fits}. The maximum temperature of the fits is 9.9~K (see Table~\ref{table:cp_fits}), and the extensions of the lines to $T^2=150~{\rm K}^2$ are extrapolations.}
\label{fig:heat_capacity}
\end{figure}

The $C_{\rm p}$ of our three samples measured in $H=0$ and in the $T$~range 1.8--300~K are presented in Fig.~\ref{fig:heat_capacity}(a). A sharp peak is evident for each sample close to its $T_{\rm N}$ as determined from the above $\chi(T)$ measurements. The $T_{\rm N}$ obtained from the $C_{\rm p}(T)$ measurements for the three compositions are listed in Table~\ref{table:cp_fits}.  The noise seen in Fig.~\ref{fig:heat_capacity}(a) for the samples with x = 0 and 0.25 is believed due to instrumental effects associated with the PPMS\@. 

\begin{table}
\caption{\label{table:cp_fits} Parameters obtained by fitting the low $T$ $C_{\rm p}\, / \, T$ versus $T^2$ data in Fig.~\ref{fig:heat_capacity}(b). Listed are: the electronic specific heat coefficient $\gamma$, the $T^3$ coefficient $\beta$, lower limits to the Debye temperature $\Theta_{\rm D}$ calculated from $\beta$ and the $T$ range of data fitted for each composition $x$. The N\'eel temperatures $T_{\rm N}$ obtained from the heat capacity measurements are also listed.}
\begin{ruledtabular}
\begin{tabular}{cccccc}
$x$		& $\gamma$ 			& $\beta$  	& $\Theta_{\rm D}$		& $T$ Range & $T_{\rm N}$  \\ 
		&	(mJ/mol K$^2$)	&	(mJ/mol K$^4$)	&	(K)	&	(K)		& (K)\\ \hline
0.00	& 24.26(7)	& 0.158(2) & 333(2) & 		1.81--9.92 & 245.8	\\
0.25	& 27.58(9)	& 0.125(2) & 360(2) & 		1.81--9.68	& 214.6 \\
0.50 	& 28.73(5)	& 0.096(2) & 393(3) & 		1.82--8.20	& 187.7 \\
\end{tabular}
\end{ruledtabular}
\end{table}

In order to isolate the electronic from other contributions to the heat capacity, the low-$T$ data are plotted as $C_{\rm p}/T$ versus $T^2$ in Fig.~\ref{fig:heat_capacity}(b).  We fitted these data by 
\be
\frac{C_{\rm p}}{T} = \gamma  + \beta T^2,
\label{Eq:C/TvsT2Fit}
\ee
where $\gamma T$ is the electronic contribution to $C_{\rm p}$, $\gamma$ is the Sommerfeld electronic heat capacity coefficient and $\beta$ is the coefficient of the $T^3$ contribution to $C_{\rm p}$. The $T$ range of each fit and the fitting parameters obtained are given in Table~\ref{table:cp_fits}. It is seen that $\gamma$ monotonically increases and $\beta$ decreases with increased Sc concentration.

The $T^3$ contribution to $C_{\rm p}$ could come from a contribution from three-dimensional AF spin waves in addition to that from the lattice.  Neglecting the possible spin-wave contribution, lower limits of the Debye temperature $\Theta_{\rm D}$ are obtained from the values of $\beta$ according to
\be
\Theta_{\rm D} = \left(\frac{12\pi^4Rn}{5\beta}\right)^{1/3},
\ee
where $n=3$ is the number of atoms per f.u.\ and $R$ is the molar gas constant.  The values of $\Theta_{\rm D}$ for the three scandium concentrations are listed in Table~\ref{table:cp_fits} and are seen to increase monotonically with $x$.

Assuming that the value of $\gamma$ is independent of $T$ below 300~K, one obtains the electronic heat capacity at $T=300$~K for each sample as $C_{\rm e}(300~{\rm K}) = \gamma T$ using the respective value of $\gamma$ in Table~\ref{table:cp_fits}.  The Dulong-Petit high-$T$ limit of the lattice heat capacity at constant volume $C_{\rm latt}$ due to acoustic lattice vibrations is $C_{\rm latt} = 3nR = 74.8$~J/mol~K\@.  The sum of $C_{\rm e}$ at 300~K and the Dulong-Petit value of $C_{\rm latt}$ is 82.1, 83.1 and 83.4~J/mol~K for $x=0$, 0.25 and 0.50, respectively, which are somewhat larger than the respective experimental values at 300~K in Fig.~\ref{fig:heat_capacity}(a).  This is consistent with expectation, since the values of $\Theta_{\rm D}$ in Table~\ref{table:cp_fits} are all greater than 300~K and therefore $C_{\rm latt}$ at 300~K for each sample is significantly below its high-$T$ limit.

\section{\label{Conclusion} Summary}

Nearly single-phase polycrystalline samples of the \lumnsi\ system with compositions $x = 0.00$, 0.25, and 0.50 were prepared and studied. Rietveld refinements of powder XRD patterns confirmed that this system crystallizes in the primitive orthorhombic TiNiSi-type structure. The lattice parameters were found to decrease almost linearly with increasing Sc content, consistent with the smaller size of Sc relative to Lu. The Mn atoms form zigzag chains. The distance between Mn atoms in these chains is found to generally decrease as the smaller Sc atoms are substituted for the Lu atoms. 

The $\rho(T)$ data for our three samples in the \lumnsi\ system showed positive temperature coefficients with magnitudes in the metallic range, indicating that this system is metallic.  The $C_{\rm p}(T)$, $\chi(T)$, and $\rho(T)$ measurements indicate the occurrence of a bulk AF transition in each sample with AF ordering temperatures $T_{\rm N} = 246$, 215 and 188~K for $x=0$, 0.025 and 0.50, respectively.  A second transition is observed at somewhat lower~$T$ for each sample from the $\chi(T)$ and $\rho(T)$ measurements at $H=0$ that is completely suppressed in $H=5.5$~T, which we speculate are due to spin reorientation transitions. 

The Sommerfeld linear heat capacity coefficient $\gamma$ and the coefficient $\beta$ of the $T^3$ contribution were extracted from the $C_{\rm p}$ data at $T<10$~K\@.  The $\gamma$ values showed enhanced values of 24.3, 27.6 and 28.7~mJ/mol~K$^2$ for $x = 0,$ 0.25 and 0.5, respectively, whereas the $\beta$ values decreased significantly with increasing~$x$ with values of 0.158, 0.125 and 0.096~mJ/mol\,K$^4$, respectively.  The $\beta$ values may include a significant contribution from three-dimensional AF spin waves in addition to the lattice contribution; with the information available we cannot separate these contributions.

We performed field- and temperature-dependent measurements of the magnetic properties up to 1000~K\@.  The high-$T$ $\chi$ data are well described by Curie-Weiss behaviors with small Mn spin $S\sim1$. The Curie constant was found to vary between the three compositions, which suggests that \lumnsi\  may be an itinerant magnetism system.  However, we proceeded to analyze the $\chi(T\leq T_{\rm N})$ data in the local moment picture as done in the past for other itinerant magnetism systems (see, e.g., Ref.~\onlinecite{Johnston2010}).

The main goal of this work was to utilize and test our recent molecular field theory for $\chi(T\leq  T_{\rm N})$ of planar noncollinear AF structures of local moments interacting by Heisenberg exchange\cite{Johnston2012} as applied to  polycrystalline samples of the \lumnsi\ system.  This system was reported from neutron diffraction measurements to have a planar helical ground state with a composition-dependent pitch angle.\cite{Venturini1997}  Analysis of our  $\chi(T\leq  T_{\rm N})$ data for three compositions of \lumnsi\  were consistent with a composition-dependent pitch angle for a helical AF ground state structure in the vicinity of either $\sim100^\circ$ or $\sim 145^\circ$; from our analysis we cannot distinguish between these two possibilities.  Within this model, we estimated the exchange interactions between a Mn spin and Mn spins in other planes of spins along the helix axis.  Due to ambiguities in the description of the AF structures in Ref.~\onlinecite{Venturini1997}, we were not able to quantitively compare our predictions for the pitch angles with the results in that reference.
 
We have shown that analyses of $\chi(T\leq T_{\rm N})$ data for polycrystalline samples, assuming a local-moment Heisenberg model for the magnetism, can reveal information about whether an AF ground state has a planar noncollinear structure, and if the noncollinear structure is a helix what the pitch of the helix is.  Further magnetic neutron diffraction measurements are needed to test our hypotheses for the composition dependence of the ground state magnetic structure of the orthorhombic \lumnsi\ system.

\acknowledgments

The work at Ames Laboratory was supported by the U.S. Department of Energy, Office of Basic Energy Sciences, Division of Materials Sciences and Engineering.  Ames Laboratory is operated for the U.S. Department of Energy by Iowa State University under Contract No.~DE-AC02-07CH11358.

\end{document}